# GIANT FREQUENCY SHIFT OF INTRAMOLECULAR VIBRATION BAND IN THE RAMAN SPECTRA OF WATER ON THE SILVER SURFACE

M.E. Kompan

*Ioffe Institute, Saint-Peterburg, Russia*
kompan@mail.ioffe.ru

*The giant frequency shift was observed in Raman spectra for inramolecular O-H vibration band. The effect was observed in SERS-condition experiment, when exciting light was focused by short-focus objective on the Ag-surface, merged in water. The shift was detected relatively to the regularl position of band, measured from the balk of water under the same other conditions.*

Water is widespread substance in our world. It play critical role in majority of processes on Earth, especially in living world. Properties of water are under study at least few centuries. Even some our natural-based measure of physical quantities, such as kilogram, are related to water. Nevertheless, new founding in the water properties appear till nowadays. For instance, not so long ago it was realised, that the "common-known" water is a mixture of two separate substances ortho- and para-water, with there own distinguishable parameters [1].

Many specific features of water originate from the role, that plays hydrogen bonds as the basis of intermolecular interaction. The arrangement of water molecules in labile structures are still the subject of study, e.g. [2,3]. The interpretation of bands in inelastic light scattering spectra for water is still the subject of discussion [4].

But, as it goes from the data of our experiment, not only the interpretations, but also the positions of bands are not completely definite.

Bellow in this paper we describe the results of experiments of Raman scattering in water. The measurements were held in back-scattering geometry, with unpolarised registration. The red (632,8 nm) excitation light of 1-2 mW power entered from upside into the cuvette, filled with pure water (Milli-Q) through thin glass (0,3 mm) window. At some part of cuvette there was a piece of silver plate, parallel to upper glass window; in the other part there was only water. The spectra were taken by micro-Raman set-up (microscope Olimpus with objective x50, spectrometer Horiba J-Y MRS 320, with the cooled CCD detector). Micro-Raman abilities were necessary to focus the light on the surface of silver plate.

Right in this last case there was detected the band, shifted from normal position. For comparison, in another type of experiment, the chamber was shifted about two millimetres, so, that through the same window the focus appeared well aside of silver plate surface, in a bulk of liquid water. The idea of experiment is illustrated at fig.1.

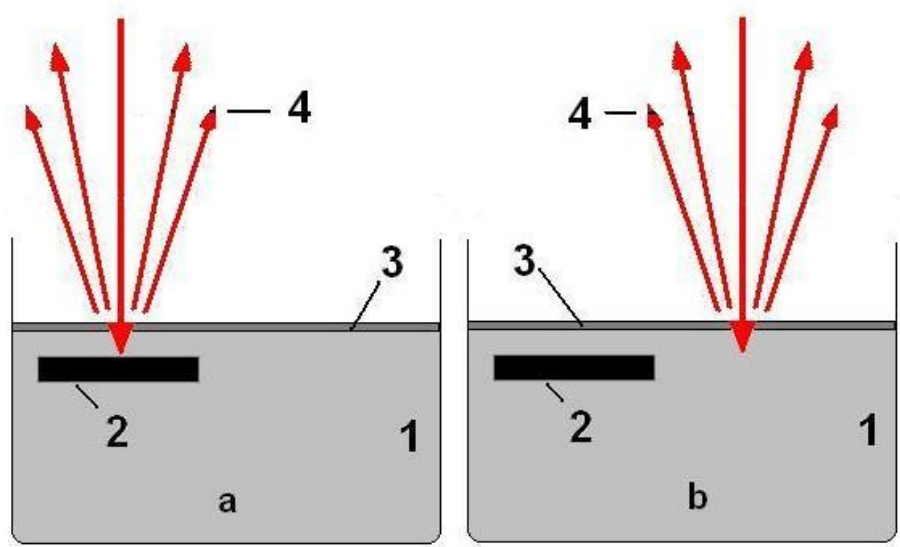

*Fig.1 a) Registration of spectra of water from Ag-surface (SERS conditions)*
*b) The cuvette is shifted, registration from bulk*
*1 – Liquid water, 2 – silver plate, 3 glass window, 4- exciting and scattered light*

The intensity of signal band from bulk water was weaker, and the resolution of subbands was worse, but the position of band (intramolecular vibration of hydroxyl in water molecule) was observed at its nominal place - $3200 \div 3400$ cm$^{-1}$.

The examples of spectra obtained are shown at fig.2,3,4. The data were subjected to minimal correction, to keep the most reliability. For instance, the subtraction of the baseline was made only in fig.4, were signal to nose ratio is worse and without this procedure the positions of band can be misread.

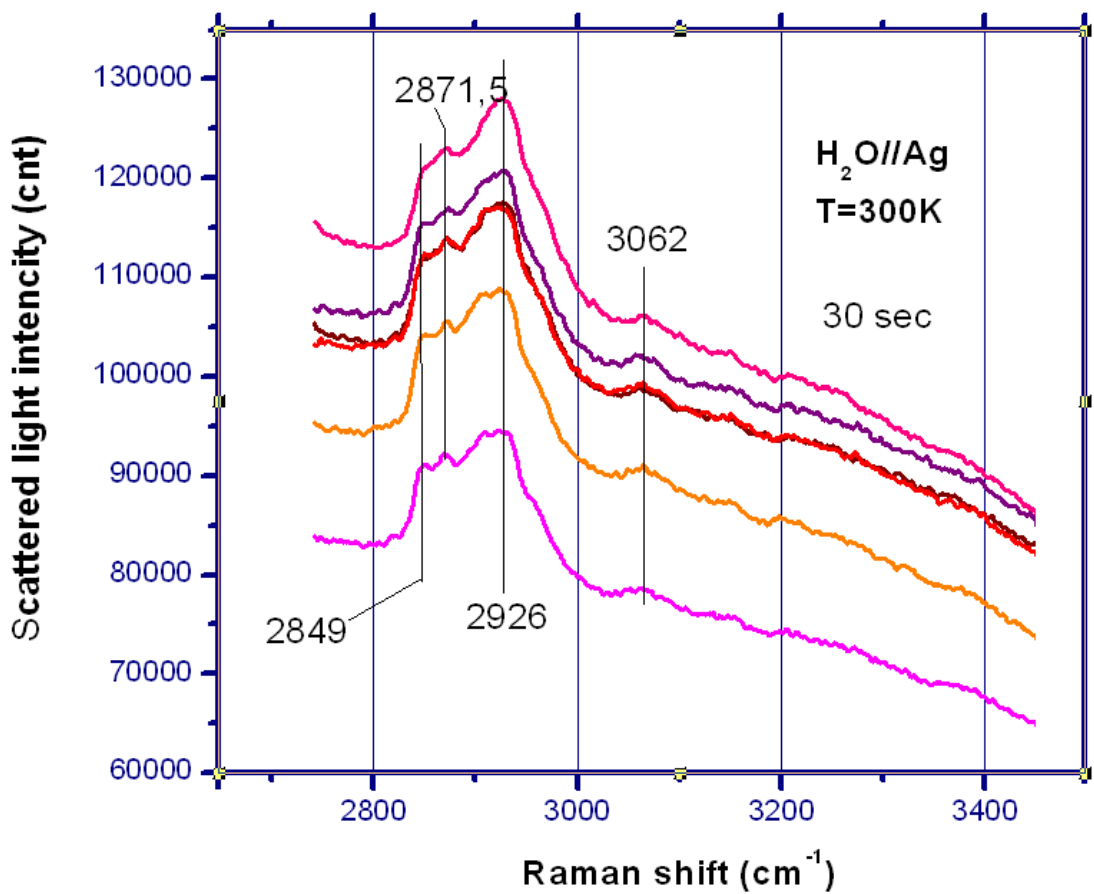

*Fig.2 Time sequence of Raman spectra from the same point on sample (SERS conditions).*

Fig.2 demonstrates the Raman spectra of water in the time sequence from the same point of silver plate. Every of six spectra were collected in 30-second acquisition. The data demonstrate the reproducibility. One can clearly observe typical for water trapezium-like band, but located in 2800-3000 $cm^{-1}$ region. A weak band is observed about 3060 $cm^{-1}$ also. No band can be find between 3200 and 3400 $cm^{-1}$ – at well know position of band of O-H intramolecular vibration.

Fig.3 allows to compare the spectra from two points on the surface. The difference between spectra is minimal. One can obviously see, that unexpected position is not due to some occasional special point on silver surface. Apart from band 3062 cm$^{-1}$ it can be seen the weak band at 3148 cm$^{-1}$. Also the traces of band can be suspected at 3200-3400 cm$^{-1}$, at usual place of water band, but they are too weak to discuss.

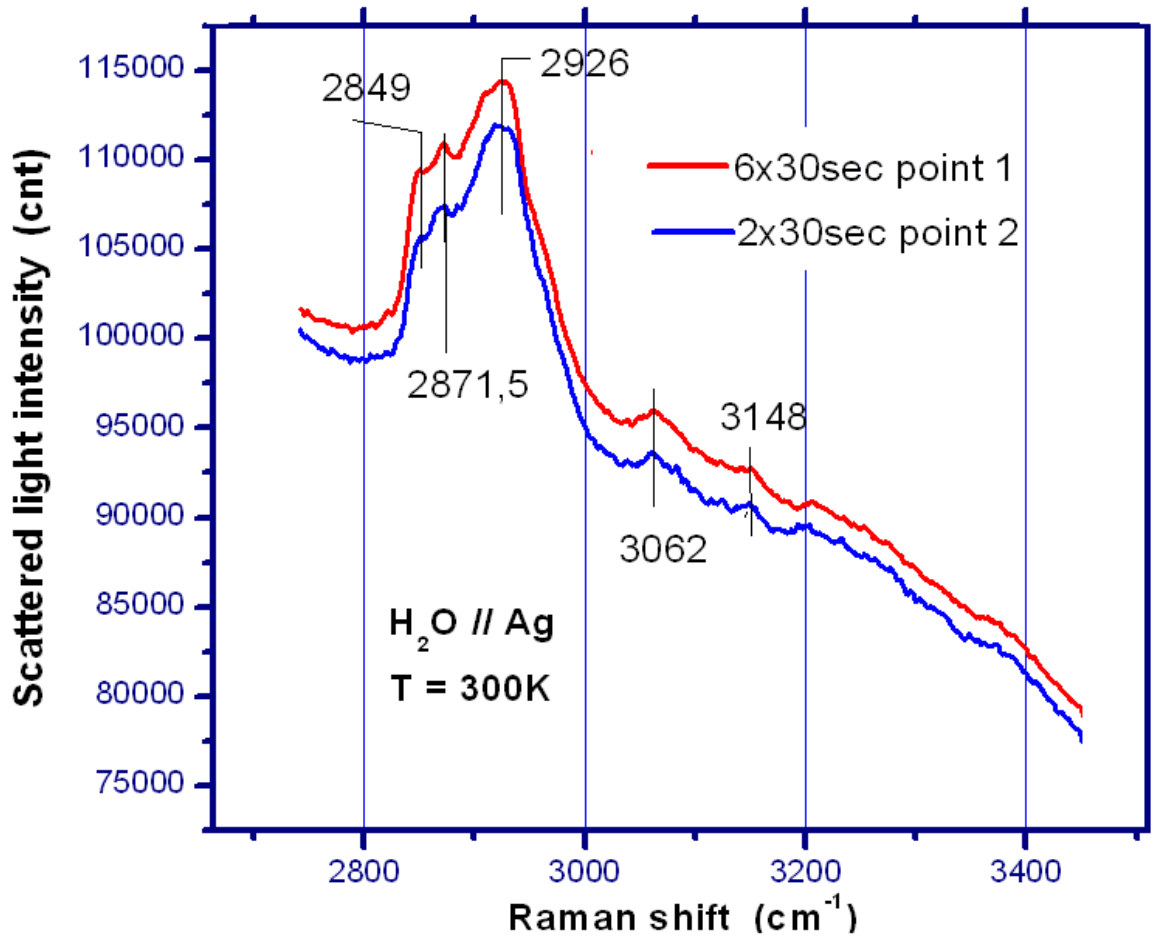

*Fig.3 Comparison of the two spectra from two different points of sample surface.*

As was mentioned, to be sure, that there weren't any experimental errors we obtained the spectra from the bulk of water. To do this, only small shift of quvette relatively optical set-up was needed. The spectra from the bulk of water is shown at fig.4. Upper blue curve is experimental spectrum without any correction. The band is observable at its "comme il faut" position. To separate the band from the slope, the baseline was subtracted. The baseline position was determined up to eye, the position and the its formula are shown at figure.

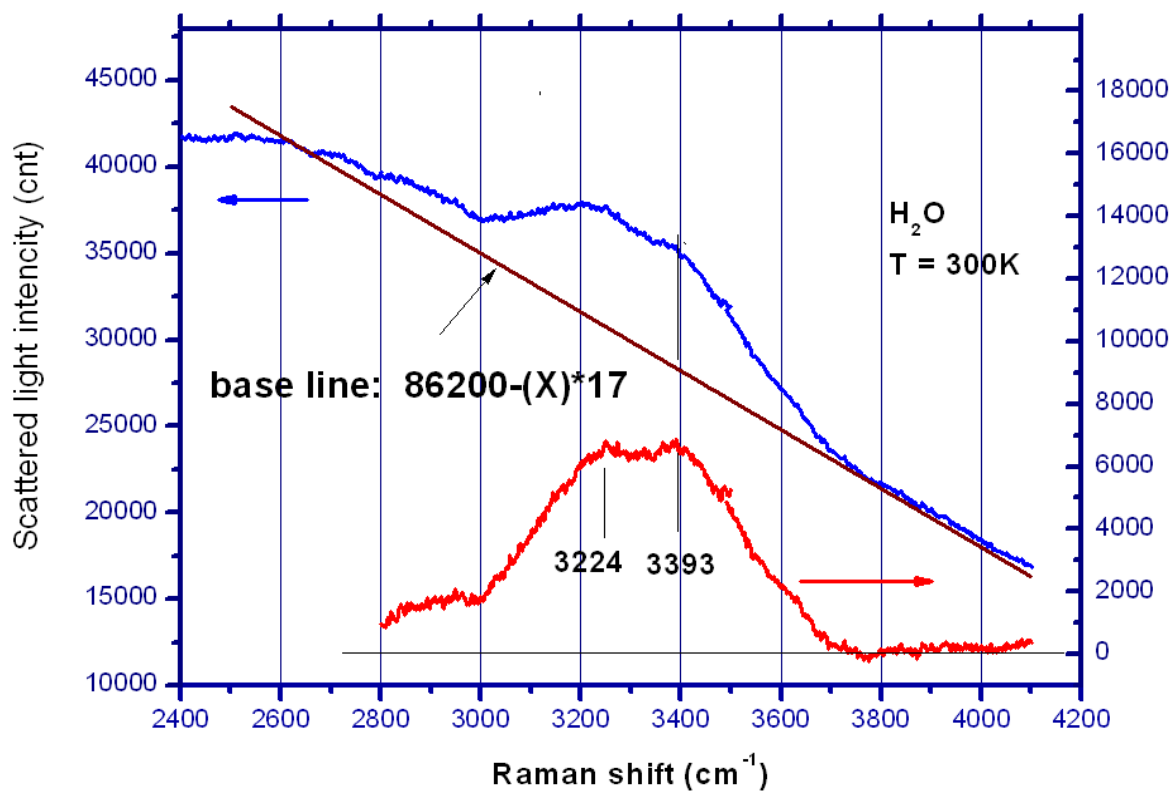

Fig.4 Spectrum from the Raman scattering in the water bulk. Blue – spectrum as obtained, red –baseline is subtracted.

So, after this comparative experiment we are sure, that the giant (more then 10%) frequency shift was detected for O-H intramolecular vibration band for water on the Ag-surface.

Also all the experimentally obtained spectra bands in 2800-3000 cm$^{-1}$ contain clearly resolvable 3-4 subbands. At this stage of a work we intentionally do not carried out the deconvolution, because this procedure isn't totally trustworthy. Now our aim is only to present the observation of giant frequency shift of spectral position of O-H vibration band. We did not find similar results published anywhere. So we consider this result, obtained in rather simple experiment, to be very interesting. At the same time it arises some questions to be answered.

First: is it really new observation? The SERS is not new, it is well known; why nobody detected so large (more then 10%) band position shift?

The answer, to our opinion, can be as follows: the most of researchers use SERS as a way to find the traces of any small additives to solution. So, maybe, that is why the researchers did not pay attention to position of water band.

Second: what is the mechanism of so large frequency shift? Is it really, that molecules of water were so much changed?

To answer this question we need remind the mechanism of SERS. The SERS effect is the result of interaction of light scattering specie (e.g. molecule) and the oscillating electrical field of free moving charge carriers in substrate (noble metals). Usually it considered, that those oscillating movement of charge carriers is induced by incident light. Our idea of explanation of the effect is that in our case the cause is in the interaction of vibrating molecule with its electrostatic image, also oscillating at the same frequency, coherently. The idea is illustrated at fig 5 on example of the dipole molecule.

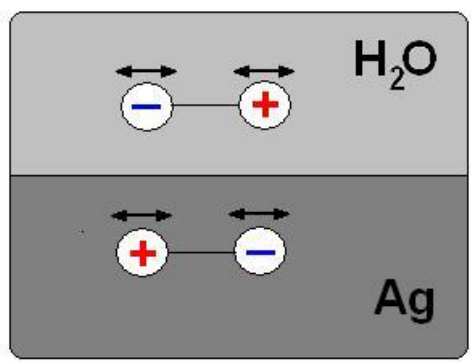

Fig.5 Mechanism of effect: coherent oscillation of mobile electrons in metal leads to decrease of interaction between ions of molecule

Let the dipole molecule vibrate somewhere near the well-conducting surface. Any charged ion (hydrogen cation or oxygen anion) induced "mirror" charge of opposite sign under the substrate surface. There must be negative "mirror" image of hydrogen and positive "mirror" image of oxygen. Those "mirror" charges situated near real ions and thus should decrease interaction between real ions. The decrease of interaction between ions in molecules has to lead to decrease in frequency of intramoleculare vibrations. That is the mechanism of the effect we observed.

Do any arguments in favour of this model can be formulated?

The intensity of observed band in spectra of water on Ag-surface is larger, then in bulk; the subbands are resolved better in this case. This is evident proof, that the light scattering is gong under SERS conditions. Thus the interaction with mobile charge carriers is necessary to take into account. Our model is based too on interaction of molecule with mobile charge carriers, it is the case.

Next: It is known that the frequency (for this band) can decrease in case of increase of a temperature. But in a second type of experiment, with a focus in a bulk of water the increase of temperature should be more, then on the surface of well-conducting metal. So, the heating cannot be the cause of the effect.

The crucial question is: whether two separate "mirror" charge images in silver can be induced? Do those induced charges are able to oscillate with necessary frequency?

Necessary to remind, that induced "mirror" charges are formed by clouds of moving electrons. The size of the induced charge is about Debye radius for the mobile charge carriers. In the macroscopic approach the time of kinetic reaction of electron clouds is about the Maxwell time. For silver, material with high electron concentration and mobility, those values are comparable with the size of water molecule and with the frequency of its vibrations. So, we think that this criteria does not contradict to our interpretation of the effect.

As a result, we suppose that our model gives correct qualitative explanation of observed effect.

So, the giant frequency shift of O-H intramolecular vibration of water in the silver surface was detected. A lot of details of the effect are left to be investigated in future. The observation of the effect can have an essential impact on our understanding of phenomena at the interfaces, and especially on understanding of phenomena on the double electrically charged layer.